# UNIVERSAL DESIGN METHODOLOGY FOR PRINTABLE MICROSTRUCTURAL MATERIALS VIA A NEW DEEP GENERATIVE LEARNING MODEL: APPLICATION TO A PIEZOCOMPOSITE

Mohammad Saber Hashemi [a], Khiem Nguyen [b], Levi Kirby [c], Xuan Song [c], Azadeh Sheidaei [d,*]

[a] Department of Aerospace Engineering, Iowa State University, 537 Bissell Road, Ames, IA 50011, United States; email: mhashemi@iastate.edu

[b] James Watt School of Engineering, University of Glasgow, Glasgow G128QQ, UK; email: khiem.nguyen@glasgow.ac.uk

[c] Department of Industrial and Systems Engineering, Iowa Technology Institute, University of Iowa, 4609 Seamans Center, Iowa City, IA 52242; email: xuan-song@uiowa.edu

[d] Department of Aerospace Engineering, Iowa State University, 537 Bissell Road, Ames, IA 50011, United States; email: sheidaei@iastate.edu

* Corresponding author

## 1.1. Abstract

We devised a general heterogeneous microstructural design methodology applied to a specific material system, elasto-electro-active piezoelectric ceramic embedded plastics, which has great potential in sensing, 5G communication, and energy harvesting. Due to the multiphysics interactions of the studied material system, we have developed an accurate and efficient FFT-based numerical method to find the multifunctional properties of diverse cellular microstructures generated by our HetMiGen code. To mine this big dataset, we used our customized physics-aware generative neural network in the format of a VAE with convolutional neural layers augmented by a vision transformer to learn long-distance features which may affect the properties of the 3D voxelized microstructures. In training, the decoder learns how to map the property distribution to the appropriate high-dimensional distribution of 3D microstructures.

Therefore, it can be considered an online material designer within the explored design space during its inference phase.

*Keywords:* Microstructure design, Discrete Fourier transform, Piezoelectricity, Ceramic–elastomer composites, Multifunctional structural optimization via generative deep learning

## 1.2. Introduction

Multifunctional microstructural materials are emerging composite or heterogeneous materials designed to possess both functional and mechanical properties dictated by their microstructures which are described by a repeating unit cell or a periodic Representative Volume Element (RVE) [1]. They can be used in a range of applications such as lightweight thermo-structural panels, blast resistant structures, high-authority morphing components [2], structural sensors [3], wearable electronics [4], and energy harvesters [5–7]. Integrating various properties and functions within a single piece of material simplifies manufacturing and reduces cost. Additive manufacturing (AM) allows precise control over materials' geometry, composition, and morphology [8–11]. While this has led to the development of materials with single properties, there has been little exploration of using microstructure to design manufacturable multifunctional materials. Designing microstructural materials with simultaneous multiple optimum properties, such as piezoelectric composites [12–14] with excellent piezoelectricity and flexibility, is a much more challenging task than designing materials based on a single property. Existing single-function design strategies suffer from several limitations, including (1) a lack of accurate and fast physical models for predicting material behavior, (2) high costs or morphological limitations of in-silico microstructure generation techniques, (3) multiple expensive offline optimization cycles for different target performances, (4) lack of modularity and generality to tackle diverse design problems, (5) reliance on purely data-driven approaches limited to the locality of convolutional operators in deep learners, (6) and imperfect translation of abstract designs into actual products due to manufacturing uncertainties and limitations. In this research effort, we aimed to develop a novel machine learning (ML)-driven design framework that can transform the current paradigm of material development into a more efficient and effective approach for designing manufacturable microstructural materials with desirable multifunctional properties. To achieve this goal, we pursued four objectives summarized in Figure 4.1: (1) developing a universal and fast in-silico voxelized microstructure generator, (2) developing accurate and rapid numerical homogenization methods for microstructural materials to obtain multifunctional and mechanical properties, (3) establishing bi-directional structure-property relationships by training our novel ML model (TransVNet) to handle sequential properties and physical constraints, and (4) validating the framework by manufacturing and characterizing designed microstructural materials using AM.

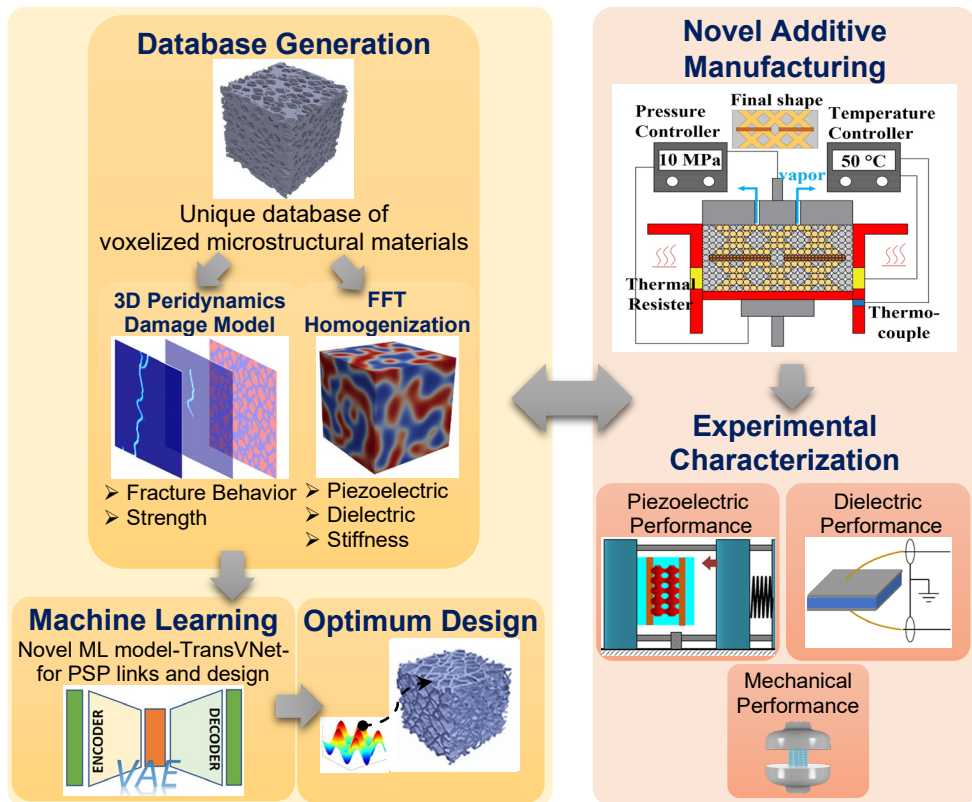

Figure 4.1. Interactive design loop of our proposed methodology for multifunctional microstructural materials.

Since microstructural material morphology influences its properties [15,16], finding the target morphology has been the focus of many studies in the field of material design. To effectively explore the design space, it is crucial to employ appropriate methods. Microstructure design methods developed based on the rich microstructure characterization and reconstruction literature [17] are limited to some target exemplars from experiments [18–20] or involve expensive simulations [21,22] and cannot be used for different material systems. For example, descriptor-based methods may be used only for particulate composites, not heterogeneous ones [23,24]. Furthermore, material design methods that focus on optimizing a high-resolution microstructure or a massive number of design variables usually require many candidate microstructures with significantly varied features to converge to the best possible solution. This, in turn, calls for a fast and efficient microstructure generation method. These two gaps in the current state-of-the-art motivated us to propose a new method for heterogeneous microstructure generation (HetMiGen) to artificially generate 3D microstructures with no need for reference 2D images of physical material. The core ideas for our proposed algorithm stem from previous studies on developing statistical microstructure reconstruction techniques for materials [25–29]. HetMiGen provides distinct capabilities not present in other synthetic microstructure generators, such as Dream3D [30]. While Dream3D is limited to producing polycrystalline microstructures, HetMiGen enables the controlled inclusion of various shapes and distributions, including anisotropy and cellular morphology. The details of our new algorithm and its computational implementation are discussed in section 4.3.

To complete the data generation for our material design methodology, an efficient computational homogenizer is needed to find effective material properties. Considering 3D voxelized microstructure models with N grid points per dimension, the computational effort for FFT simulations scales with $N \log N$, which is much more efficient than the $N^3$ scaling observed in FEM simulations. Therefore, we have developed FFT simulation codes to numerically solve the governing PDE equations of the homogenization problem. There has been a strong interest in developing FFT computational methods for multifunctional and multi-phase composite materials. However, for piezoelectricity, the methods available in the literature [31,32] are not suitable to converge to the solution in the heterogeneous microstructures whose phase boundaries are complex and irregular, and the property contrast ratio of composite constituents is high. Also, the enthalpy formulation found in the previous FFT work [31,32] leads to a saddle-point variational minimization problem which is more challenging to solve than the convex variational minimization via the energy formulation of the constitutive laws. Our homogenization method is explained in section 4.4.

Computational material design has emerged as a promising approach to accelerate the discovery of new materials with tailored properties and functions. Traditionally, material knowledge has been explored using physics-based Process-Structure-Properties (PSP) models and/or experiments in the forward direction (from process to structure to properties). However, material design inherently drives discovery in the inverse direction, which requires the establishment of reduced-order PSP linkages [33,34] using regression or statistical learning techniques on data produced by PSP models and physical experiments. Alternatively, novel mining techniques may be used on the big data generated [22,35–37] through accurate and efficient computational models with limited experimental calibrations and validations. The second route is more suitable for material systems in which the process is directly linked to the material's structure via advanced additive manufacturing [38,39] in mesoscale or traditional computer numerical control (CNC) machining of a topology-optimized (TO) structure in macro-scale [40–42]. However, TO [43] methods lack the generalizability of the newer computational methods based on generative models in ML [22,36,37] since they solve the optimization problem for each user-specified set of goals and constraints. In contrast, the ML-based ones are trained once and then used in the inference mode almost instantly with much less computational cost. At the same time, TOs are more computationally demanding even for one target property due to their dependency on FEM. Although purely convolutional neural networks (CNNs) have become widespread in ML-based methods, they lack an understanding of the input image's long-range relationships due to the intrinsic locality of convolution operators, resulting in a weak performance for structures with a large inter-sample variation. That is where Transformers [44] came into play by leveraging their self-attention mechanisms. However, their naive use may lead to poor results as they focus exclusively on the global contexts of 1D sequences. Recent research has demonstrated superior performance by combining Vision Transformers (ViTs) [45] with CNN feature extractors [46]. Building upon these advancements, our network (TransVNet) directly works with 3D voxelized microstructures instead of stacks of 2D slices and is most beneficial when the quantities of interest are affected by long-range voxel/phase relationships that are not captured by the local convolutional operators used in CNNs. It will also be useful when those quantities are sequential and inputs are in a mixed format, which are challenging to encode using current material design methods. The network has been developed such that physics and context constraints can be considered invariances in training, following the rise of physics-informed networks [47] in scientific machine learning. This will make the computational framework more efficient and modular for more complex future applications while avoiding using data-hungry deep learners commonly used in computational materials science. The network and its relation to material design is discussed in section 4.5.

Polymer-ceramic composites have tremendous potential for multifunctional devices, such as energy harvesting devices (e.g., wearable electronics and healthcare sensors) [48] and high-frequency antennas [49], as they come with tunable and advantageous mechanical properties such as high compliances, toughness, and yield strains. However, the polymer-ceramic composites with randomly dispersed ceramic particles have poor piezoelectric properties [50] (Figure 4.2(a)). Furthermore, the polymer-ceramic composite of 1–3 has superb energy harvesting properties only in the fiber direction (Figure 4.2(b)). To overcome these challenges, the mesoscale geometry of composites should be optimized by tailoring the ceramic phase morphologies. In recent years, there has been an increased interest in developing 3–3 (Figure 4.2(c)) complex geometry of composites, for example, 3D interconnected composites with continuous ceramic skeletons that possess superb energy harvesting [51,52]. Also, this material system is suitable for high-frequency antenna applications (such as 5G), and the morphology significantly affects the electric permittivity and loss in different frequency ranges. A considerable challenge of fabricating such composites is to precisely control the complicated mesoscale geometry featured by the randomness and interconnectivity of the interfaces and achieve the desired functional properties in the final structures. With the merits of dealing with complex geometries, additive manufacturing has been widely studied and applied to fabricate polymer-ceramic composites [52,53]. There have been a few efforts to numerically generate this kind of geometries [54,55]. However, such constructed geometries can still not guarantee fiber morphology and interconnectivity [54]. A few studies have been devoted to the topological optimization of piezoelectric structures [56–58]; however, our proposed design methodology does not have the conventional restriction of the TO methods mentioned above. With these insights, we have systematically investigated the effect of mesoscale geometry on stiffness, piezoelectric, and dielectric properties of flexible polymer-ferroelectric composites. Furthermore, we designed these material systems for target performances using our new deep generative network. This network has been trained on the dataset comprising various cellular morphologies generated through our HetMiGen algorithm and their corresponding homogenized multiphysics properties obtained using our high-fidelity/accurate fast FFT method. The details of manufacturing and experimental characterization are discussed in section 4.6. To show the capabilities of our proposed methodology framework, we have run our computational codes and validated them based on the PDMS-BTO material system with PDMS/matrix and BTO/inclusion properties obtained from [59,60] and [61], respectively. We experimentally measured the piezoelectric and dielectric properties of multiple samples synthesized based on a microstructure in our in-silico generated dataset to validate our computational results.

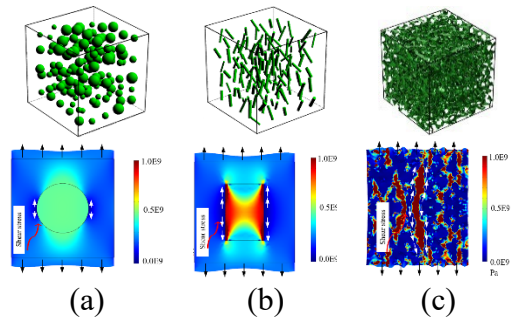

(a)          (b)          (c)

Figure 4.2. Effect of geometry on the shear stress distribution in polymer-PZT composites, (a) 0–3 particles; (b)1–3 cylinders; (c) 3–3 interconnected [50].

### 1.3. HetMiGen generating a novel virtual database of cellular microstructures

A fast and computationally efficient program, called HetMiGen (short for Heterogeneous Microstructure Generator), has been developed to artificially generate 3D heterogeneous microstructures without the need to any reference 2D cut section images of physical material microstructures through expensive experimental methods such as SEM. This in-silico data generation is especially useful for computational materials science since it does not rely on the reconstruction techniques based on some expensive experiments, let alone the physically material processing complications with the shortcoming of not being able to consider all the design space for material design purposes. The C++ source codes can be compiled for and deployed on different machines with Linux or Windows OSs, and the executable can be readily run given a CSV file whose each line contains the microstructural parameters of the microstructure to be generated: the microstructure id number, the number of phases in addition to the background phase, the volume fraction of each phase, the number of initial seeds, the increment (positive or negative) of seeds' addition in the next seed addition iterations, the frequency by which seeds are added, the radius of the local neighborhood to be checked for proximity for each phase (zero means no check is needed and seeds can be grown until they touch others), whether each phase should be clustered at the end, the rate of growth decay for each phase (zero means that growth parameters are fixed throughout the evolution iterations), and the probabilistic growth thresholds for Cellular Automate [62] based on the considered neighborhood type (von Neumann in the current version of the codes). By changing the input parameters, it can generate multitudes of heterogenous microstructures of different material systems consisting of two or more material phases with different morphologies. The parameters have affinity to the physical process of manufacturing and thermomechanical evolution of microstructure as well.

Although the program is efficient in its serial implementation (e.g., generating a bincontinous $150^3$ resolution microstructure in around two minutes normally), for massive data generation in ML/data-driven material science, the codes have been written using parallel programming techniques of OpenMP for single HPC nodes or workstations (shared memory architectures) and OpenMPI for multiple HPC nodes or cloud computing servers (heterogenous memory architectures). Due to loop-carried dependencies and to minimize communication overheads, task parallelization through SPMD style of programming was implemented to gain speed-up by dividing microstructure generation tasks into a pool of processors. To demonstrate the speed-up gain, 32 identical input data with the same seeding of random generators were considered so that the computational loads are balanced among all processors and between different configurations (from 16 cores to one core). A decent speed-up in each step was achieved as shown in Figure 4.3 for the OpenMP version. The imperfection may be attributed to the thread management (forking and joining) and memory contention on the CPU cache level.

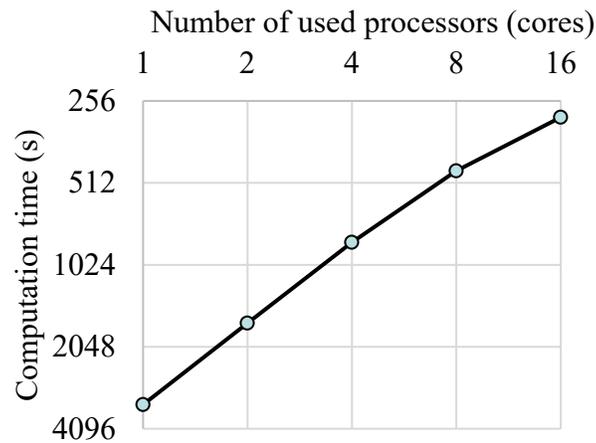

Figure 4.3. Speed-up for different configurations in a log-log scale.

As shown in the algorithm flowchart of Figure 4.4, HetMiGen algorithm is based on two basic mechanisms and an addendum to create the special type of microstructures with clustered phases such as bincontinous or cellular ones: Seed Addition, Seed Growth, and Clustering. During Seed Addition, if the current iteration is designated for such operations according to the frequency of seed addition, at most as many as the current number of seed addition, which is the summation of the previous number of seeds and the increment parameter, are dispersed in the cuboid domain of the microstructure according to the quasirandom Sobol sequence [63] for uniform dispersion in 3D space. Proximity Check1 means that the chosen seed position which should be an empty voxel (encoded with the background color) can turn into a phase voxel (encoded with a pre-designated foreground color) if its neighbors in a r-radius neighborhood, with r being one of the input parameters, are all empty. In the top right 2D image of the figure, the white pixels are the current empty or background ones, the dark blue ones are the seeds which have been added before the current stage of Seed Growth for that image, and the light blue ones are the pixels grown according to cellular automata. Therefore, there is an iterative process of seed addition in each iteration of the microstructure evolution until all current number of seeds are added, or the process proves to be stagnant following an exhaustive search for the right position of the last seed. Immediately after this part, the cells will be grown probabilistically in different directions according to a cellular automata algorithm and the chosen neighborhood type like the von Neumann one shown by the cross-hatched pixels of the 2D image. The thresholds for the probabilistic growths are calculated as the decay factor times the initial threshold, and the random trial for each neighborhood voxel is simulated by the pseudo-random number generator MT19937 (Mersenne Twister) [64] resulting in a float number between 0 and 1 drawn from a uniform distribution in every function call. So larger thresholds in each direction mean higher chances of growth in that direction leading to different levels and directions of anisotropy in the final evolved microstructure. In Seed Growth, Next interface voxel is a phase voxel which has at least one empty voxel in its immediate von Neumann neighborhood, which can be considered as a potential location to perform Cellular Automata. Proximity Check2 is similar to Proximity Check1 except that the empty condition is appended by an alternative: if the voxels in its local neighborhood all belong to the same cluster of phase voxels (called a cell hereafter). After each iteration of Seed Growth, the volume fraction of each foreground phase is calculated. If they are close to the target input parameters within an acceptable tolerance, the program will be terminated by converting the binary data of the last evolution of the microstructure into 8-bit BMP images, as shown by the post-processed/smoothed image of the 3D bone microstructure in the figure. Another alternative for termination is that if the evolution has been stagnant, i.e., there has been no increase in current volume fractions after 5 evolution iterations, the program is failed, and it will attempt to regenerate by starting the random sequences (Sobol for seed positioning and MT19937 for seed growth) with another seeding number. If the failure persists after five trials, the program will move on to generate the next requested microstructure.

If a foreground phase should be clustered according to the input parameters, the volume fraction check will be done based on a lower volume fraction target (85%) to let the clustering algorithm work with current cells by connecting them and adding more foreground phase voxels. The clustering algorithm tries to connect the isolated cells of the grown seeds where they are close to each other by passing through all empty voxels, checking whether at least two isolated cells of the material phase to be clustered are present in the local r-radius neighborhood of the current empty voxel, and turning all the empty voxels in the neighborhood into the phase color if at least two isolated cells are found, and the measured phase density in the local neighborhood is higher than the threshold in the current iteration. This process is also iterative in that the phase density threshold is incrementally decreased to prioritize connections where the cells are closest to each other, and if the density threshold cannot be further lowered (i.e., zero), the radius of the local neighborhood for density measurement is increased. The iterative algorithm is terminated if the volume fraction goes out of a tolerance band around the requested volume fraction, or if more than 90% of the phase voxels are connected to each other.

As mentioned in the introduction, the design space of heterogeneous or cellular microstructures is huge as each voxel of $150^3$-resolution 3D image can be attributed to a single material phase. Since our design methodology is based on machine learning to find the complex bi-directional structure-property mappings, we generated a big dataset of bincontinous or cellular microstructures, represented by their 2D slice images, to consider different morphological possibilities and to train our network on a highly diverse and inclusive dataset. Therefore, a list of different HetMiGen input parameters, including different volume fraction numbers and anisotropy and dispersion/seed numbers, was generated according to Sobol sequence within ranges specified by the physical constraints and sensitivity analysis in the code results. As illustrated in Figure 4.5, the resulting dataset was skewed despite the maximum dispersion nature of Sobol sampling because some sets of input parameters are very challenging, i.e., taking too much computational time, or even impossible to be successfully simulated. Therefore, a fixed number of microstructures was considered in each volume fraction bin to have a balanced exposure in the network training.

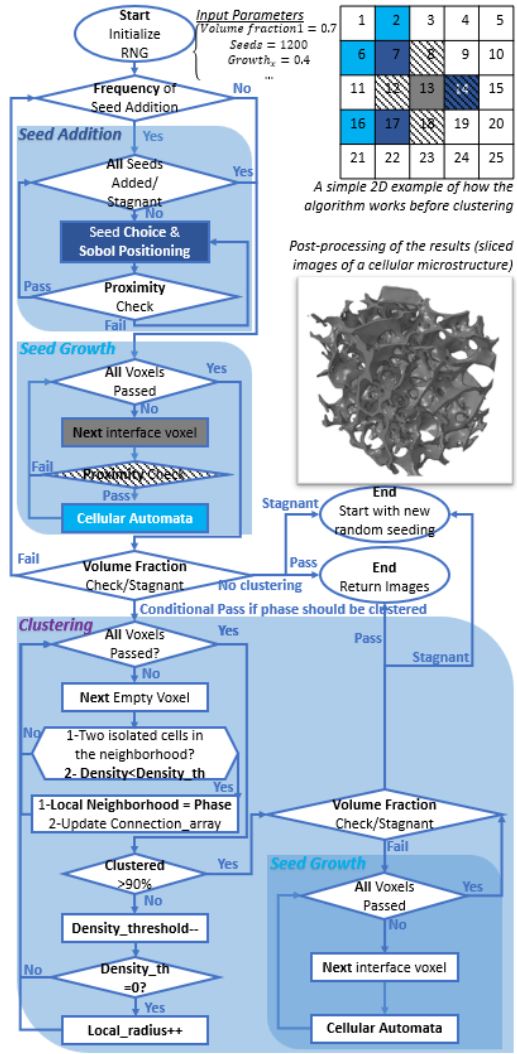

Figure 4.4. Graphical abstract of the algorithm implemented in our HetMiGen program.

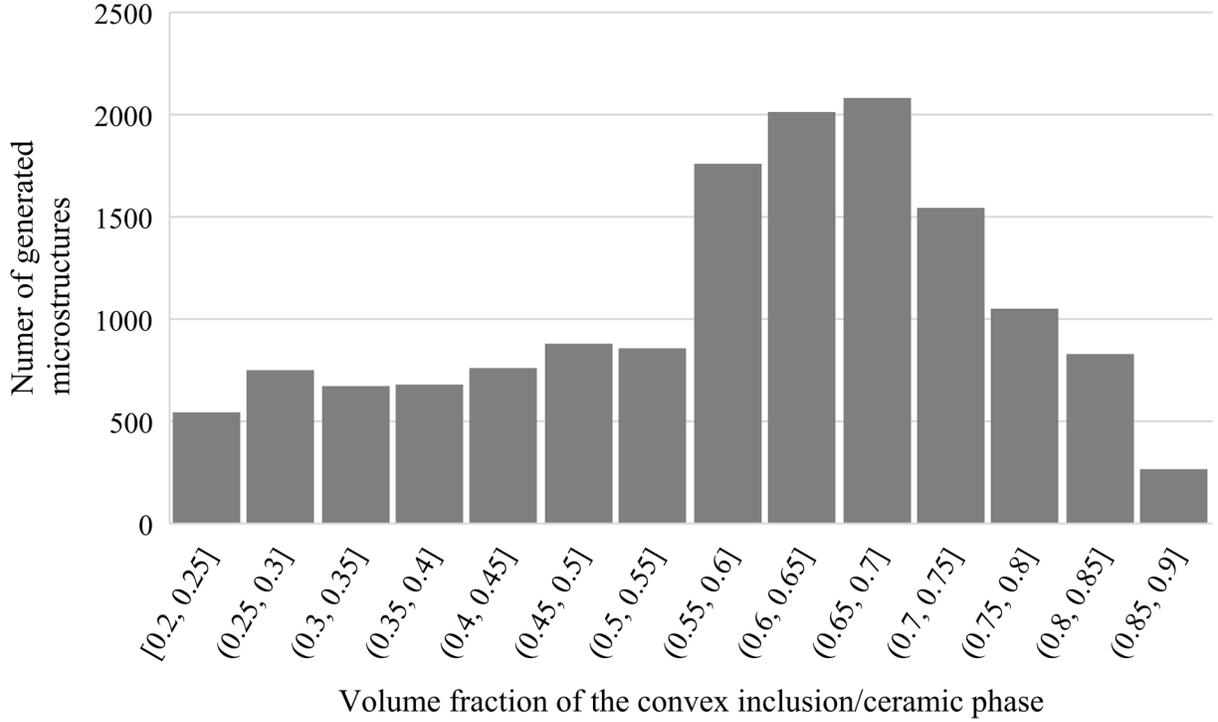

Figure 4.5. Histogram of the microstructures' volume fraction (VF) distribution.

### 1.4.    Characterizing the Generated Microstructures via our Multiphysics FFT Homogenization Method

Applications of piezoelectric materials in small strain regime is the focus of this study. The deformation and the electrostatic status of an electro-mechanical continuum body can be described by the displacement $u$, a vector field, and the electric field potential $\phi$, a scalar field. The gradient fields induced by them are the linearized strain field $\epsilon = \frac{1}{2}(\nabla u + u\nabla)$ with its energetic conjugate of stress tensor field $\sigma$ and the electric vector field $E = \nabla\phi$ with its energetic conjugate of electric displacement vector field $D$, respectively. The quasi-static mechanical governing PDE equations are the compatibility one, $\nabla \times \epsilon \times \nabla = 0$, and the balance of linear momentum in the absence of external forces, $\nabla . \sigma = 0$. There are four possible constitutive formulations according to the desired input and output variables: Stress-Charge, Stress-Voltage,

Strain-Charge, and Strain-Voltage. Only strain loading is considered in macroscale in this paper, and these are the corresponding linear constitutive relations:

Table 4.1. Linear electroelastic constitutive relationships.

| Stress-Charge form | Stress-Voltage form |
|---|---|
| $\sigma = \mathbb{C}_{E=0} : \epsilon - e^T . E$ $D = e : \epsilon + \chi_{\sigma=0} . E$ (4.1) | $\sigma = \mathbb{C}_{D=0} : \epsilon - q^T . D$ $E = -q : \epsilon + \chi_{\sigma=0}^{-1} . D$ (4.2) |

The fourth-order tensors $\mathbb{C}_{E=0}$ and $\mathbb{C}_{D=0}$ are the elastic stiffness of the material in zero electric and charge fields, respectively, the second-order tensor $\chi_{\sigma=0}$ is the electric permittivity in zero mechanical stress, and the third-order tensors $e$ and $q$ are, respectively, the piezoelectric coupling coefficients for the Stress-Charge and Stress-Voltage forms, respectively. The Stress-Voltage form can be derived as the partial derivatives of the energy density function

$$W(\epsilon, D) = \frac{1}{2}\epsilon : \mathbb{C}_{D=0} : \epsilon - D . q : \epsilon + \frac{1}{2}D . \chi_{\sigma=0}^{-1} . D : \qquad \sigma = \frac{\partial W}{\partial \epsilon}, E = \frac{\partial W}{\partial D}. \qquad (4.3)$$

Similarly, by carrying out the partial Legendre transformation, we reach at the enthalpy density function

$$W^*(\epsilon, D) = inf_E[W(\epsilon, D) - D . E] = \frac{1}{2}\epsilon : \mathbb{C}_{E=0} : \epsilon - E . e : \epsilon + \frac{1}{2}E . \chi_{\sigma=0} . E \qquad (4.4)$$

where

$$\mathbb{C}_{D=0} = \mathbb{C}_{E=0} - e^T . \chi_{\sigma=0}^{-1} . e; \qquad q = \chi_{\sigma=0}^{-1} . e. \qquad (4.5)$$

As a result, the Stress-Voltage form can be derived by identifying

$$\sigma = \frac{\partial W^*}{\partial \epsilon}; \qquad D^* := -D = \frac{\partial W^*}{\partial E}. \qquad (4.6)$$

From now on, $\mathbb{C} := \mathbb{C}_{D=0}, \mathbb{C}^* := \mathbb{C}_{E=0}, \chi := \chi_{\sigma=0}$. The problem of material characterization by homogenization can be described as finding appropriate fields' functions that minimize the total energy or enthalpy functional of an RVE or a cell such that they observe the periodic boundary conditions, the mechanical compatibility equation, and Gauss's law or Faraday's law, and the averaging conditions. Specifically, an RVE occupying the 3D domain $\mathcal{R} \subset \mathbb{R}^3$ consists of $N_{mat}$ material phases characterized by $W^{(p)} = W^{(p)}(\epsilon, D)$ and $W^{*(p)} = W^{*(p)}(\epsilon, E)$ each occupying a subdomain $\mathcal{R}^{(p)} \subset \mathcal{R}$. Therefore, the electromechancial energy density of the cell is

$$W\big(x, \epsilon(x), D(x)\big) = \sum_{p=1}^{N_{mat}} \zeta^{(p)}(x) W^{(p)}\big(x, \epsilon(x), D(x)\big) \ \forall x \in \mathcal{R}, \tag{4.7}$$

where $\zeta^{(p)} = \zeta^{(p)}(x)$ is the phase indicator function which maps each position vector to value 1 in $\mathcal{R}^{(p)}$ and 0 otherwise. Considering the volume average of a quantity as $\langle(\cdot)\rangle = \frac{1}{\mathcal{R}} \int_{\mathcal{R}} (\cdot) dV$, and assuming the cell is subject to the macroscopic strain and the macroscopic electric displacement or the macroscopic electric field, the averaging conditions are

$$\langle\epsilon\rangle = \bar{\epsilon}; \langle D\rangle = \overline{D}; \langle E\rangle = \overline{E}. \tag{4.8}$$

Following the derivation of Lippman-Schwinger equations and the Green operators $\Gamma$, $L$, and $G$, the equilibrium equations are as follows. For the minimization problem of the energy formulation, the equations in the real space are

$$\Gamma * \sigma = 0, L * E = 0 \Leftrightarrow \Gamma_{ijkl} * \sigma_{kl} = 0, L_{ij} * E_j = 0. \tag{4.9}$$

In the Fourier's space, they transform into

$$\hat{\Gamma}: \hat{\sigma} = 0, \hat{L}.\hat{E} = 0 \Leftrightarrow \hat{\Gamma}_{ijkl}\hat{\sigma}_{kl} = 0, \hat{L}_{ij}\hat{E}_j = 0. \tag{4.10}$$

In parallel, for the saddle point problem of the enthalpy formulation, their counterparts are

$$\Gamma * \sigma = 0, G * D = 0 \Leftrightarrow \Gamma_{ijkl} * \sigma_{kl} = 0, G_{ij} * D_j = 0;$$

$$\hat{\Gamma}: \hat{\sigma} = 0, \hat{G}.\hat{D} = 0 \Leftrightarrow \hat{\Gamma}_{ijkl}\hat{\sigma}_{kl} = 0, \hat{G}_{ij}D_j = 0. \tag{4.11}$$

In the energy formulation, if we split the strain field and the electric displacement vector into the sum of their average or macroscopic fields and the fluctuation ones as $\epsilon = \bar{\epsilon} + \tilde{\epsilon}, D = \overline{D} + \widetilde{D}$, the aforementioned equilibrium equations lead to

$$\Gamma * \left( \mathbb{C} : \tilde{\epsilon} - q^T . \widetilde{D} \right) = -\Gamma * \left( \mathbb{C} : \bar{\epsilon} - q^T . \overline{D} \right);$$

$$L * \left( -q . \tilde{\epsilon} + \chi^{-1} . \widetilde{D} \right) = -L * \left( -q . \bar{\epsilon} + \chi^{-1} . \overline{D} \right). \tag{4.12}$$

This linear system of equations for the solution fields $\tilde{\epsilon}$ and $\widetilde{D}$ must be supplemented with the averaging conditions $\langle \tilde{\epsilon} \rangle = 0$ and $\langle \widetilde{D} \rangle = 0$. The right-hand sides are known given the applied macroscopic boundary conditions or the input arguments of the solution algorithm, $\bar{\epsilon}$ and $\overline{D}$. By applying Fourier transforms and then inverse Fourier transforms, we have an equivalent system of equations

$$\mathcal{F}^{-1}\{\hat{\Gamma} : \mathcal{F}[\mathbb{C} : \tilde{\epsilon} - q^T . \widetilde{D}]\} = -\mathcal{F}^{-1}\{\hat{\Gamma} : \mathcal{F}[\mathbb{C} : \bar{\epsilon} - q^T . \overline{D}]\};$$

$$-\mathcal{F}^{-1}\{\hat{L} : \mathcal{F}[-q . \tilde{\epsilon} + \chi^{-1} . \widetilde{D}]\} = -\mathcal{F}^{-1}\{\hat{L} : \mathcal{F}[-q . \bar{\epsilon} + \chi^{-1} . \overline{D}]\}, \tag{4.13}$$

that can be solved by a conjugate-gradient solver since the left-hand side can be thought of as the effect of the linear operators of $\tilde{\epsilon}, \widetilde{D}$ and the right-hand side is a given computable vector. Consequently, the system can be viewed as $Ax = b$, where $Ax$ represents the entire linear operator and is computed in a matrix-free fashion. We consider $\frac{\|b - Ax\|}{\|b\|} < TOL$ as the stopping criterion of the iterative solver with TOL given by the user. Similarly, the corresponding Fourier equations for the enthalpy formulation are

$$\mathcal{F}^{-1}\{\hat{\Gamma} : \mathcal{F}[\mathbb{C}^* : \tilde{\epsilon} - e^T . \widetilde{E}]\} = -\mathcal{F}^{-1}\{\hat{\Gamma} : \mathcal{F}[\mathbb{C}^* : \bar{\epsilon} - e^T . \overline{E}]\};$$

$$-\mathcal{F}^{-1}\{\hat{G} : \mathcal{F}[-e . \tilde{\epsilon} - \chi . \widetilde{E}]\} = -\mathcal{F}^{-1}\{\hat{G} : \mathcal{F}[-e . \bar{\epsilon} - \chi . \overline{E}]\}, \tag{4.14}$$

Note that equation $G * D^* = -G * D = 0$ is solved instead of the original $G * D = 0$ so that the left-hand side has a symmetric matrix $A$. Nevertheless, this is resulted from the saddle-point variational problem, and it cannot be solved by standard CG solver easily. To this end, we used the Minimum Residual Method to solve the system for the enthalpy formulation.

Considering the symmetries in our derivation, the number of required operations to solve the Fourier equations and find the homogenized properties is almost halved compared with conventional methods described in the literature. That is especially crucial in the real-world applicability of our data-driven design framework by having the most robust numerical homogenizer for material characterization. Following the selection of microstructures in the dataset discussed in the previous section, we performed FFT homogenization simulations given their images (geometrical material phase information) to find their linear electromechanical properties. For each microstructure, nine different far-field unit load cases were solved to find the homogenized composite property matrices since the linear property matrix is nine by nine in the most general case. The orthotropic constituent properties, in the Stress-Charge form, are calculated as follows:

$$\begin{bmatrix} \sigma_{11} \\ \sigma_{22} \\ \sigma_{33} \\ \sigma_{23} \\ \sigma_{13} \\ \sigma_{12} \\ D_1 \\ D_2 \\ D_3 \end{bmatrix} = \begin{bmatrix} C_{11} & C_{12} & C_{13} & 0 & 0 & 0 & 0 & 0 & -e_{31} \\ C_{12} & C_{11} & C_{13} & 0 & 0 & 0 & 0 & 0 & -e_{31} \\ C_{13} & C_{13} & C_{33} & 0 & 0 & 0 & 0 & 0 & -e_{33} \\ 0 & 0 & 0 & C_{44} & 0 & 0 & 0 & -e_{15} & 0 \\ 0 & 0 & 0 & 0 & C_{44} & 0 & -e_{15} & 0 & 0 \\ 0 & 0 & 0 & 0 & 0 & C_{66} & 0 & 0 & 0 \\ 0 & 0 & 0 & 0 & e_{15} & 0 & \gamma_{11} & 0 & 0 \\ 0 & 0 & 0 & e_{15} & 0 & 0 & 0 & \gamma_{11} & 0 \\ e_{31} & e_{31} & e_{33} & 0 & 0 & 0 & 0 & 0 & \gamma_{33} \end{bmatrix} \cdot \begin{bmatrix} \varepsilon_{11} \\ \varepsilon_{22} \\ \varepsilon_{33} \\ \varepsilon_{23} \\ \varepsilon_{13} \\ \varepsilon_{12} \\ E_1 \\ E_2 \\ E_3 \end{bmatrix}.$$

(4.15)

The composite property matrices were then post-processed such that only the 11 prominent components, as defined in the orthotropic property matrix above, were extracted as the continuous label vectors for each microstructure. The Figure 4.6 results show a high level of variation in the calculated properties as the range of variation with respect to the average values is between 0.02 and 6.38. To facilitate the ML training process, these label vectors were normalized such that the average and the variance of each property component is zero and one, respectively.

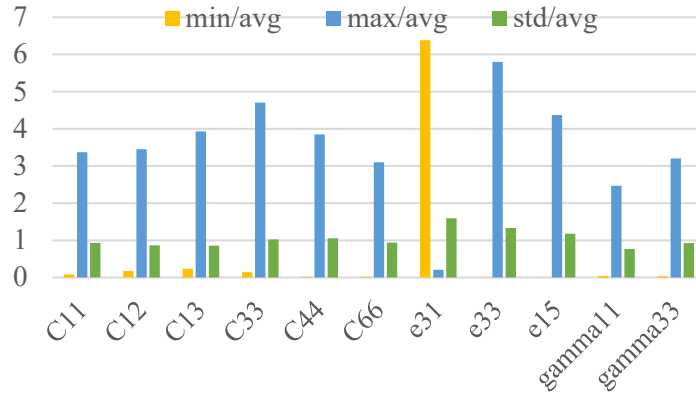

Figure 4.6. The range of homogenized properties variation with respect to their average values in the homogenized microstructures.

## 1.5. TransVNet, a New Robust ML Model to Find SP Links

The cornerstone of our design methodology is our new machine-learning framework called TransVNet. At a high level, the goal of microstructure optimization is achieved by creating two nonlinear maps: (a) the forward problem, which maps a given microstructure (e.g., a two-phase microstructure) to a specific set of properties, and (b) the inverse problem which, for a given target set of properties, produces a microstructure with the desired properties. However, both the forward problem, which typically requires solving coupled nonlinear partial differential equations (PDEs), and the inverse problem, which involves solving a PDE-constrained nonlinear optimization problem, are computationally challenging. The current morphology optimization approaches are heavily reliant on multiple forward-model PDE calculations. Therefore, there is an urgent need for ML approaches that can efficiently and accurately solve both the forward and inverse problems in a single setting. Our proposed TransVNet network (illustrated in Figure 7) addresses this challenge. We aim to approximate the forward and inverse maps using deep generative models, a new type of neural network widely used in content generation, image translation, and style transfer, among other applications [65–67]. However, their application in engineering design, materials design (especially heterogeneous microstructural design), and scientific computing has not been explored to the extent of their capabilities in literature.

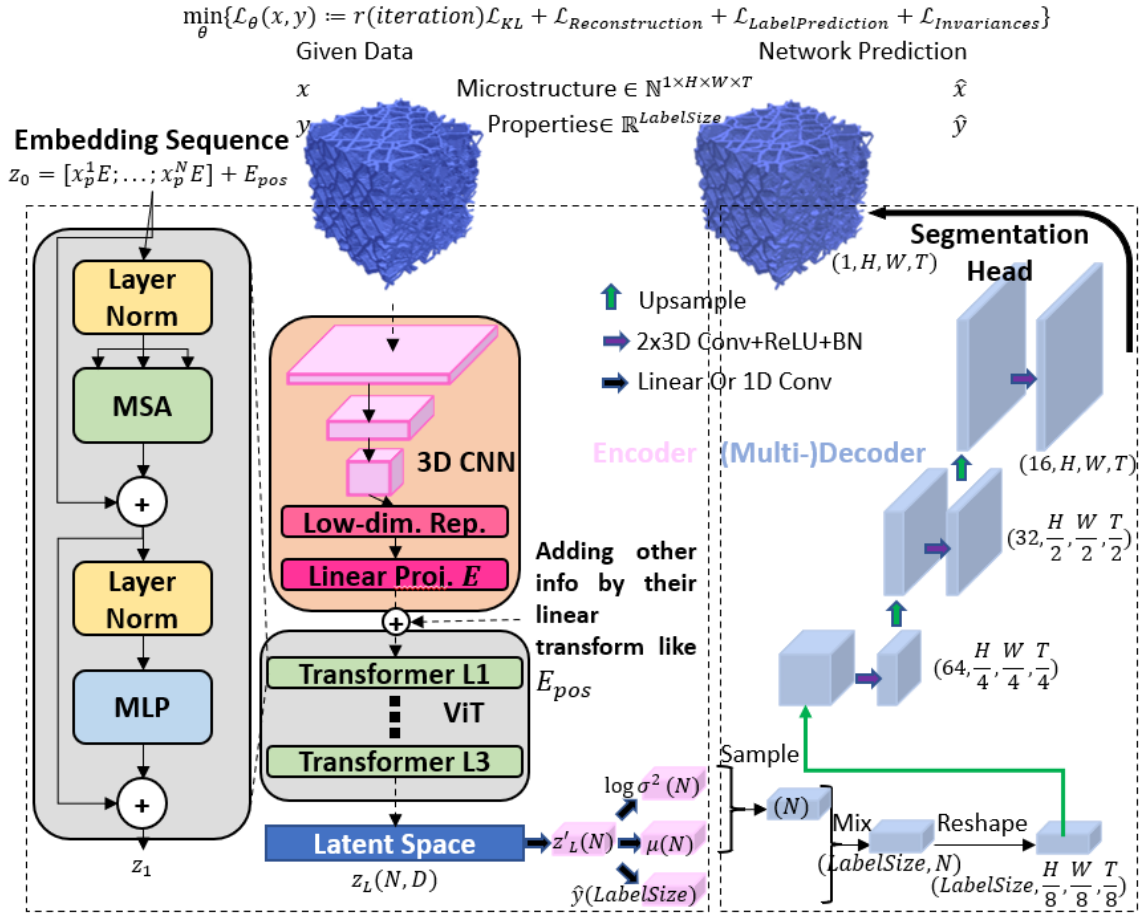

Figure 4.7. Schematic diagram of our proposed ML method, TransVNet, for material design.

Our proposed TransVNet, inspired by TransUNet [46], is suitable for 3D image and mixed-format data, which is the case for the current study as the best way to work with heterogeneous microstructures is to work with their images, not just limited expert-picked features such as its auto-correlation statistics. We also did not choose to use TransUNet or similar 2D networks on the stack of 2D images as it cannot fully exploit the 3D relationships between different volumetric microstructural regions. Furthermore, our new TransVNet will be semi-supervised in that its decoder will be used as a generative model for material design following the training of the whole network consisting of its encoder and its decoder. To the best of authors' knowledge, this is the first generative model for the material design that leverages the state-of-the-art ViTs in computer vision. Other notable differences are our customized CNN feature extractor in the encoder instead of pre-trained CNNs (such as ResNet [68] as used in TransUNet [69]) since there is no pre-trained CNN for 3D images, its capability to include expert knowledge in the embeddings fed to the ViT which can improve the performance of the network and minimize the training costs, and considering a composite loss function since the network is a variational autoencoder for the purpose of material design. Given an image $x \in \mathbb{N}^{C \times H \times W \times T}$ with C, H, W, and T being the number of information channels (it is one in this study because each voxel represents the material phase at that location, which is itself a categorical variable), the height, the width, and the thickness of the 3D image (all 64 following the resizing of our HetMiGen $150^3$-resolution microstructures) and $y \in \mathbb{R}^{LabelSize}$ with $LabelSize$ being the number of features of a label vector, the goal is to classify each voxel with a label (a material phase in this study) on the decoder side. Following a CNN feature extraction in the encoder resulting in a low-dimensional representation of the original 3D input image, with $C'(=32 \ in \ this \ study)$ channels and a dimension size of $H' = \frac{H}{2^{\#CNNDownScaling}} = \frac{64}{2^3} \ (in \ this \ study)$, and the method of Dosovitskiy et al. [45], the input image is tokenized into a sequence of $N$ flattened vectors of 3D image patches of size $P(= 1 \ in \ this \ study)^3$; with $N = \frac{H'W'T'}{P^3}$: $\{x_p^i \in \mathbb{R}^{(P^3.C')} | i = 1, \dots, N\}$. The vectorized patches $x_p^i$ are then projected into a D($= 72$ in this study)-dimensional embedding space by learnable linear transformation, $E \in \mathbb{R}^{(P^3.C') \times D}$. To retain the spatial information and to include the salient microstructural features from expert knowledge if it proves to be needed for better performances or cost-saving, positional embedding, $E_{pos} \in \mathbb{R}^{N \times D}$, are added to the patch embeddings as follows $z_0 = [x_p^1 E; x_p^2 E; \ldots; x_p^N E] + E_{pos}$. The Transformer encoder is composed of $L(= 3 \text{ in this study})$ layers of Multi-head Self-Attention (MSA) and Multi-Layer Perceptron (MLP) blocks such that the output of layer $l$, $z_l$, is the following transformation applied to the previous layer output, $z_{l-1}$:

$$z_l' = MSA\big(LN(z_{l-1})\big) + z_{l-1};$$

$$z_l = MLP\big(LN(z_{l'})\big) + z_{l'}, \tag{4.16}$$

where $LN(.)$ is the layer normalization operator. To avoid information loss in low-level features such as the shape and the boundary of the material phases, and to improve the accuracy of the network in inference, a hybrid CNN-Transformer architecture constitutes the encoder. The output of the last Transformer's layer or $z_L \in \mathbb{R}^{N \times D}$ is convoluted into a single channel vector whose each component represent the Transformer's input token ($z_L' \in \mathbb{R}^N$), and it is then transformed by decoupled fully connected perceptron layers to find the latent distribution parameters ($\mu = z_L' FC_{avg}$; $FC_{avg} \in \mathbb{R}^{N \times N}$ for the average parameters and $\log(\sigma^2) = z_L' FC_{\log\_var}$; $FC_{\log\_var} \in \mathbb{R}^{N \times N}$ for the log of variance parameters) and the network-predicted labels ($\hat{y} = z_L' FC_{labels}$; $FC_{labels} \in \mathbb{R}^{N \times LabelSize(=11 \text{ in this study})}$ ). Our latent information mixing is unconventional as we consider each property component or label feature as the attenuation or bias parameter for a single information channel of the final latent variable fed into the decoder as the generative model. Following the method of Kingma et al. [70], we have the following for training:

$$z \sim q(z|x) = \mathcal{N}(\mu, \sigma^2); \big(decoder_{input}\big)_{ij} = z_j + \hat{y}_i \in \mathbb{R}^{(i \in LabelSize) \times (j \in N)}, \tag{4.17}$$

where $z \sim q(z|x)$ means z has a probability distribution conditioned on $x$, and $\mathcal{N}$ denotes the normal distribution. In the inference, $z$ is sampled from our desired latent distribution, $p(x) = p(x|\theta)p(\theta) = \mathcal{N}(x|0, I)$ with $I$ being the identity matrix, and $\left(decoder_{input}\right)_{ij} = z_j + y_i$ with $y$ being the desired label, i.e., the target properties in this study. Considering $\theta$ as the trainable parameters of our variational autoencoder, the composite loss function to be minimized in training is as follows:

$$\min_{\theta}\{\mathcal{L}(\theta) := r(iteration)\mathcal{L}_{KL} + \mathcal{L}_{Reconstruction} + \mathcal{L}_{LabelPrediction} + \mathcal{L}_{Invariances}\};$$

$$\mathcal{L}_{KL} = E_{q_{\theta}}[\log(q_{\theta}(z|x)) - \log(p(x))];$$

$$\mathcal{L}_{Reconstruction} = -\log\left(p(x|y,z)\right) = -\log\left(p(x|\hat{x} \sim \mathcal{N}\left(decoder_{output}(y,z), \bar{\sigma}^2\right))\right), \bar{\sigma}$$
$$= \theta_I, i.e., trainable;$$

$$\mathcal{L}_{LabelPrediction} = MSE(y, \widehat{y_{\theta}}) = \frac{1}{n}\sum_{i=1}^{n}\left(y_i - \widehat{y_{l_{\theta}}}(x)\right)^2,$$

(4.18)

where $\mathcal{L}_{KL}$ is the Monte-Carlo estimation of the KL divergence or the closeness measure between the network-achieved latent distribution, $q_{\theta}(z|x)$, and the target isotropic normal distribution, $p(x)$, the negation in $\mathcal{L}_{Reconstruction}$ guides the optimizer such that the network generates images which are similar to the given inputs by maximizing the likelihood of observing $x$ given $\hat{x}$, and $\mathcal{L}_{Invariances}$ is generally considered for future physics-informed learning purposes (it is 0 in this study). $r(iteration)$ is the cyclical annealing schedule, a cyclical linear function of the training iteration in this study, which has been shown to improve the training stability and final performance of VAEs by allowing the progressive learning of more meaningful latent codes and leveraging the informative representations of previous cycles as warm re-starts [71]. The $decoder_{input}$ is reshaped, from $LabelSize \times N$-size tensor to $LabelSize \times \frac{H'}{P} \times \frac{W'}{P} \times \frac{T'}{P}$, to recover the shape of an image for further upscaling. Multiple upsampling steps are done to decode $decoder_{input}$ and output a newly generated 3D image in inference or reconstruct the input image in training. The cascaded upsampler consists of decoder blocks with a trilinear upsampler succeeded by two ReLU activated convolutional layers. Following some experiments, a single CUP led to poor performance on some test cases related to vastly different volume fraction ranges. Therefore, the final network has seven CUP heads dedicated to different volume fraction ranges to improve the general network performance.

The generated dataset was split into two sets: training dataset consisting of 85% of samples and testing or evaluation dataset. The training results in Figure 4.8 show that the loss values have decreased through back-propagation optimization via ADAM optimizer such that they have reached an almost stationary state. The fluctuations are due to the cyclic multiplier of the KL divergence loss as discussed above. This multiplier, taking values between zero and one, is displayed inside the figure as a function of the number of training epochs. In this training trial, the number of epochs is 200, and each epoch takes 9 iterations to go over the whole training dataset. To assess the performance of the trained model, the unseen test dataset was used. The labels, i.e., property vectors, were given to the decoder alongside a random sample of the isotropic normal distribution. The generated images were then fed into the trained encoder as the fast surrogate of the FFT homogenizer to estimate the properties of the designed microstructures and compare them with the values obtained from FFT homogenization in the test dataset. The regression plots for two important properties according to the application of the studied material system (flexible sensors) were drawn to assess the level of accuracy of the surrogate model, shown in Figure 4.9(a) for the C33 stiffness property and in Figure 4.9(b) for the e33 piezoelectric property, as well as the material designer, shown in Figure 4.9(a) for the C33 stiffness property and in Figure 4.9(b) for the e33 piezoelectric property. The ideal result associated with maximum accuracy is when all data points are located on $y = x$ line. Therefore, it can be inferred from the plots and $R^2$ values that the models are generally accurate except where the underlying property manifold has not been explored equally in our generated dataset, i.e., the high-volume fraction regions. Despite this lower level of accuracy, the models are still capable of finding the solutions qualitatively.

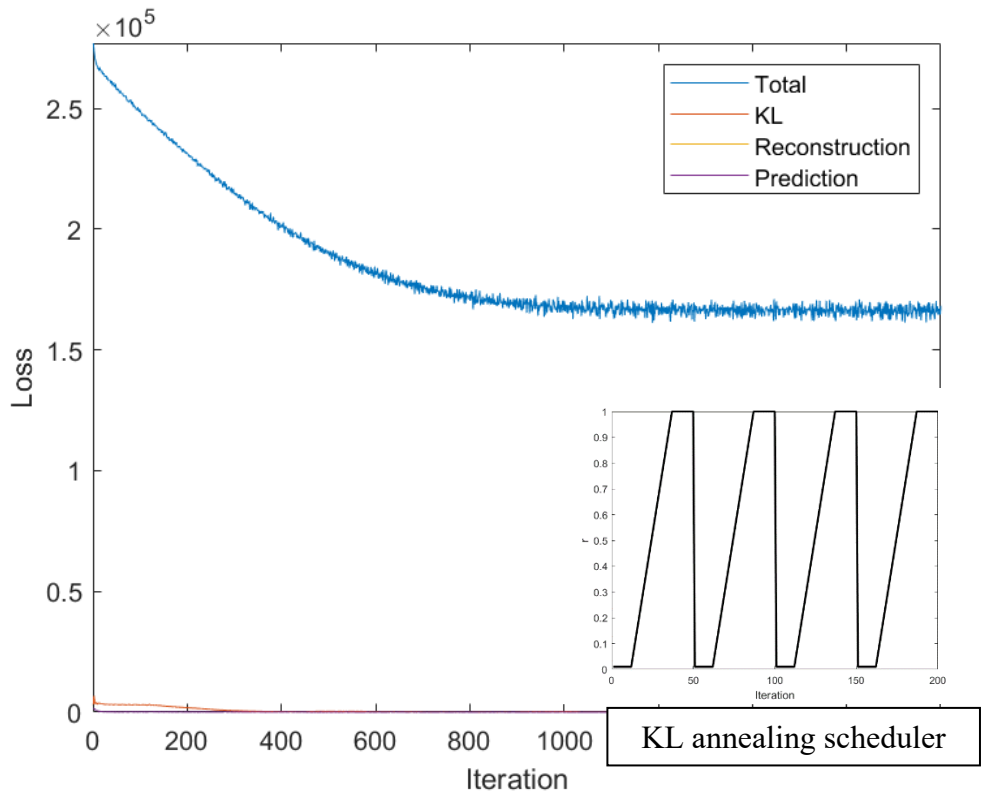

Figure 4.8. The training loss values and KL annealing scheduler.

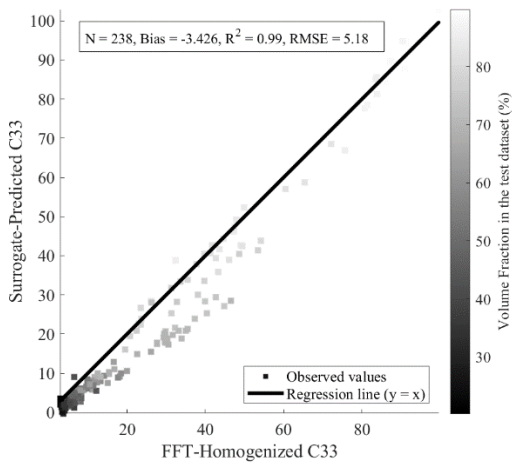

(a)

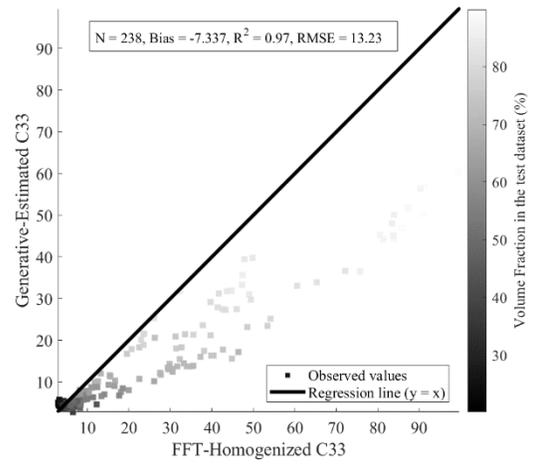

(b)

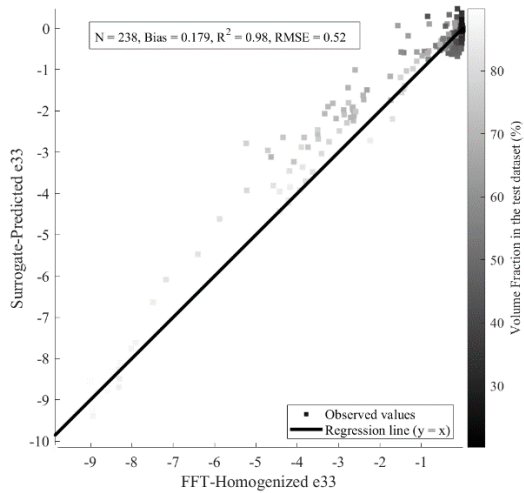

(c)

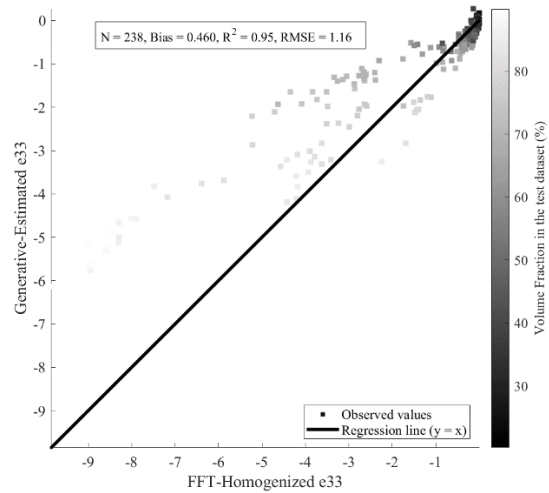

(d)

Figure 4.9. The performance of the trained models shown by their regression plots with FFT-Homogenized properties as the ground truth values on x axes (C33 unit: $GPa$; e33 unit: $C/m^2$).

## 1.6. Experimental validation

Collaboration with The University of Iowa's AMPRL Lab provided the fabrication process for the tested samples, based on a process from part literature [72]. A patented support-free ceramic AM technology, i.e., suspension-enclosing projection stereolithography [73] was utilized to fabricate piezoelectric ceramic-polymer composites.

The feedstock material for printing consisted of a ceramic powder, a combination of photocurable resins, and a dispersant. For exceptional dielectric properties, barium titanate (BaTiO3, < 2 μm, Sigma-Aldrich, Darmstadt, Germany) was selected as the model ceramic material at a concentration of 70 wt%. Two commercial resins were selected with desirable photosensitivity at concentrations of 15 wt%. Formlabs resin (FLGPCL01, Formlabs, Boston, USA) and 15 wt% Anycubic resin (white, ANYCUBIC, Shenzhen, China). To unify the particle distribution, Solsperse 20000 (Lubrizol, United Kingdom, Europe) was chosen as the dispersant at 1 wt% of the final composition. Materials were manually mixed to visible uniformity and then ran in a planetary ball mill for 2 hours at 300 RPM to ensure the slurry was truly homogeneous. For final preparations, the slurry was degassed in vacuum at -1.5 bar for 30 minutes.

In short, the process enables the support-free fabrication of piezoceramic components with complex overhanging structures by harnessing an elasto-viscoplastic ceramic suspension feedstock and a meniscus-backflow-based layer-recoating method. A schematic of the printing process is shown in Figure 4.10, provided by Li et. al [73]. The suspension contains photopolymer binder which is cured in a layerwise fashion by a UV projector on the stereolithography process. An image of a projected layer is provided in Figure 4.11(a) which comes from a sliced layer of a CAD model. The projector not only cures the image in the suspension in the x-y direction, but also penetrates in the z-direction to bond each layer together. Following the printing process, the UV-cured structure is cleaned of any excess suspension, as shown in Figure 4.11(b). Once printed, the green composite contains significant amounts of photopolymer resin. Using a tube furnace (STF150, Carbolite-Gero LLC, Hope Valley, UK), the printed part was fired at $600°C$ for 3 hours under Argon to decompose the resin. The residue char remained to bond the BTO particles. Following debinding, the parts were placed in a regular muffle furnace (HTF18, Carbolite-Gero LLC, Hope Valley, UK) at 1330 °C for 240 min. Once finished sintering, the final part was a fully dense and organic ceramic. Sintered parts were taped on one end to ensure polymer did not coat the testing surface. Following, they were placed in an FDM printed cubic vessel. Polydimethylsiloxane (PDMS), an elastomer, was directed into the vessel using a syringe until the polymer reached the top surface of the ceramic, without breaching the surface. The FDM vessel was placed in vacuum for 10 minutes to ensure the polymer had reached all cavities within the complex structure. The sample during the infiltration process and once removed from the vessel is shown in Figure 4.11(c). Excess polymer was cut from the ceramic presenting the final part tested (Figure 4.11(d)).

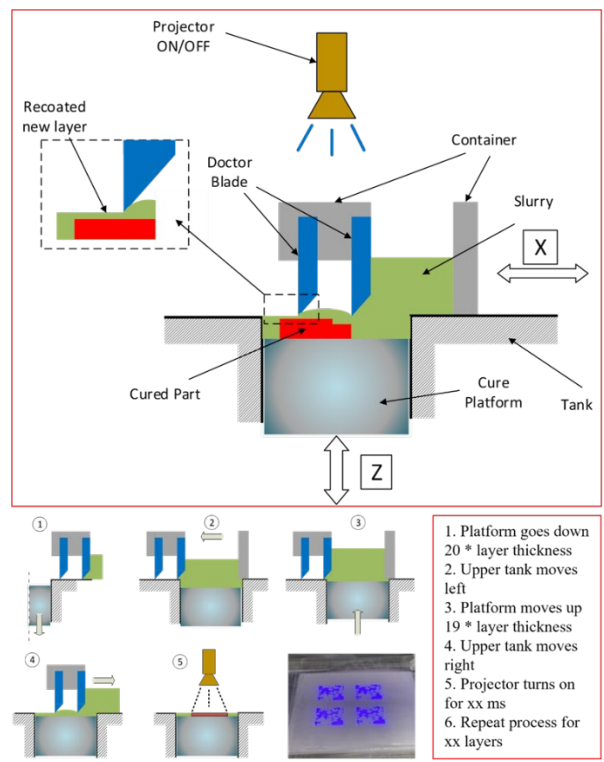

Figure 4.10. The printing process schematic [73].

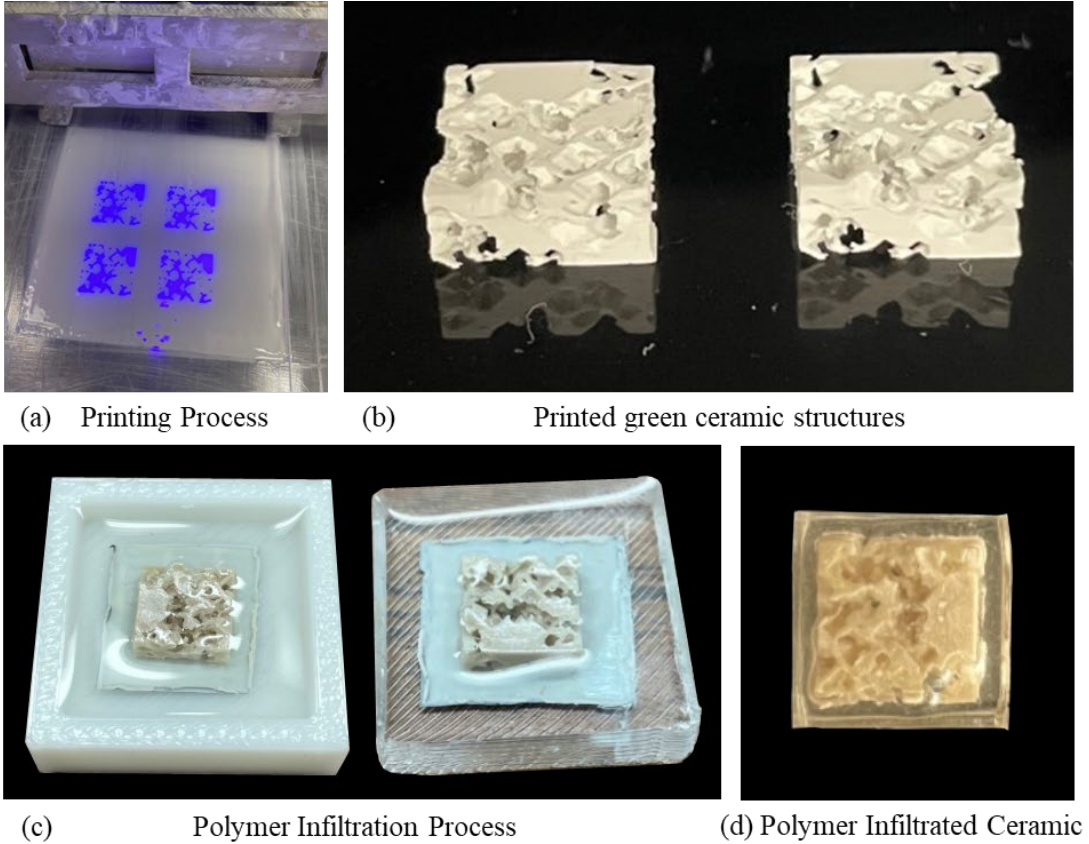

(a)  Printing Process     (b)     Printed green ceramic structures

(c)     Polymer Infiltration Process     (d) Polymer Infiltrated Ceramic

Figure 4.11. The process for polymer infiltrated ceramics include (a) printing, (b) green structures, (c) polymer infiltration, and (d) the final part.

In this research, $d_{33}$ was measured using a piezoelectric charge constant meter (PKD3-2000, PolyK Technologies LLC, Philipsburg, USA) at a force-frequency of 110 Hz. The piezoelectric voltage constant was calculated via the equation $g_{33} = d_{33}/\varepsilon_r\varepsilon_0$. Dielectric Permittivity was calculated via $\varepsilon_{33} = \varepsilon_r\varepsilon_0$, where $\varepsilon_0 = 8.854 \times 10^{-12} \ F/m$ is the permittivity of air, and $\varepsilon_r$ is the relative permittivity of the measured material [72]. The $\varepsilon_r$ and $\tan\delta$ were measured using an LCR meter (TG2811D, TONGHUI, Changzhou China) at 1 kHz frequency, signal source output resistance of 100$\Omega$, a signal level of 1.0 V, and a parallel equivalent circuit. For preparation, the samples were lightly coated on the top and bottom surface with Nickel conductive paint (MG Chemicals, Ontario Canada) to ensure maximum conductivity. The relative permeability and loss tangent for all 4 samples were calculated. The experimental results are shown in Table 4.2. The FFT homogenization of the CAD model from which the experimental samples were synthesized is as follows: $(\varepsilon_r)_{33} = \frac{\gamma_{33}}{\varepsilon_0} = 250.14; d_{33} = e_{33}(C^{-1})_{33} = 122.3 \frac{pC}{N}$. Therefore, our homogenization method resulted in satisfactory estimation of the physical properties based on the average measurements.

Table 4.2. The experimental measurements and comparison with our computational model.

| Sample | 1 | 2 | 3 | 4 | Average | Computational |
|---|---|---|---|---|---|---|
| Relative Permittivity $(\varepsilon_r)_{33}$ | 315.39 | 239.07 | 293.44 | 292.04 | 284.985 | 250.14 |
| Loss Tangent | 0.011 | 0.01165 | 0.0099 | 0.01075 | 0.01083 | NA |
| Piezoelectric Charge Constant $d_{33}$ ($\frac{pC}{N}$) | 112 | 73 | 53 | 74 | 78 | 122.3 |

## 1.7. Conclusions

In this study, we have proposed a universal methodology to design multifunctional heterogeneous or microstructural materials based on several novel computational methods validated with the experimental results for the specific application of flexible composites of ceramic-embedded plastics. Our first method was HetMiGen to realize different geometrical possibilities for heterogeneous microstructures, especially the cellular or bincontinous ones with superior properties in flexible piezocomposites, without any need to experimental data. Its results indicate its versatility and computational efficiency using C++ parallel programming. We selected a balanced dataset from its results for optimum training. We developed our efficient FFT homogenizer to find the effective composite properties of the selected dataset. Its results showed the vast design possibilities by the level of variation in each property component. The cornerstone of our methodology is our 3D and long-range inferring deep learner, called

TransVNet, based on the state-of-the-arts in the computer vision field as well as our customized architecture for its physics-aware applications, including our current static design problem. The network is a variational autoencoder so that its trained encoder acts as the fast surrogate of costly numerical properties homogenizer, such as the FFT one in this study, and its trained decoder performs the generative design process given target properties, which is challenging due to the very large number of design variables ($64^3$ in our trial) and the complexity of the multiphysics properties manifold as each material phase contributes to a specific desired property in the composite. Nevertheless, our results show high accuracy of the surrogate model as well as satisfactory performance of the material designer. This material designer is also advantageous compared with other conventional methods since it can be constantly improved by providing more data in the future through transfer learning and used almost instantaneously or online in contrast with optimization-based methods. As seen in the regression plots for the case study, the property manifold has not been explored in balance, and the designer accuracy may be improved by generating more diverse set of high-volume fraction microstructures. The current study is computational yet purely data-driven resulting in the valuable byproduct of in-silico generated dataset of different material microstructures labeled with their multifunctional properties as well as the trained network, all expandable in future studies for other material systems and applications. Also, our computational framework has been built considering the future problems such as context-aware learning, e.g., the physical and manufacturing constraints as terms of the loss function, and dynamic or sequential property learning, e.g., the electrical properties as functions of frequency. Lastly, we presented the experimental details of the additive manufacturing technique used to make the studied material system. Although the results point to a strong link between our computational framework and the real-world application, future studies should focus on decreasing the mismatch between these two and increasing the repeatability with lower variations of the composite properties among synthesized samples of a given microstructure geometry.

**Data Statement**

The code and the dataset are made available on GitHub and can be found in

https://github.com/ms-hashemi/HetMiGen.

**Funding**

This work was supported by Iowa State University.